\documentclass[conference]{IEEEtran}
\usepackage{graphbox}
\IEEEoverridecommandlockouts

\usepackage{amsmath,amssymb,amsfonts}
\usepackage[ruled, linesnumbered, noend]{algorithm2e}
\usepackage{graphicx}
\usepackage{rotating} 
\usepackage{booktabs}
\usepackage{multirow}
\usepackage{makecell}
\usepackage{textcomp}
\usepackage{xcolor}
\usepackage{xspace}
\usepackage[
    colorlinks=true,
    citecolor={blue!70!black},
    linkcolor={red!60!black},
    urlcolor={blue!70!black},
    breaklinks=true
]{hyperref}

\everydisplay{\small}
\newcommand{\inlinemathsize}{\footnotesize}
\SetAlCapFnt{\small}
\SetAlFnt{\small}
\SetAlCapNameFnt{\small}
\usepackage[
  backend=biber,
  maxbibnames=1,
  minbibnames=1,
  hyperref=true,
  backref=false
]{biblatex}
\renewbibmacro{in:}{}

\def\BibTeX{{\rm B\kern-.05em{\sc i\kern-.025em b}\kern-.08em T\kern-.1667em\lower.7ex\hbox{E}\kern-.125emX}}

\newcommand{\projtitle}{Maestro\xspace}

\newcommand{\cpara}[1]{\vspace{0.3em}\noindent\textbf{#1}}

\begin{document}

\title{\projtitle: Workload-Aware Cross-Cluster Scheduling for LLM-Based Multi-Agent Systems}

\author{
    \IEEEauthorblockN{Jinghao Wang\textsuperscript{1}, 
    Xiao Zhou\textsuperscript{1},
    Xiaoyang Sun\textsuperscript{2\dag},
    Yihui Zhang\textsuperscript{1}, 
    Yilong Li\textsuperscript{1},
    Tianyu Wo\textsuperscript{1}, 
    Xu Wang\textsuperscript{1}, \\
    Chunming Hu\textsuperscript{1},
    Renyu Yang\textsuperscript{1}
    \thanks{\dag  Dr. Xiaoyang Sun is the corresponding author}}
    \IEEEauthorblockA{
    \textsuperscript{1}Beihang University \quad\quad  \textsuperscript{2}University of Leeds \\
    \{wang\_jinghao, zhouxiao2021, zhangyihui, liyiloong, woty, xuwang, hucm, renyuyang\}@buaa.edu.cn; x.sun4@leeds.ac.uk
    }
}

\maketitle

\begin{abstract}
Large Language Model based Multi-Agent Systems (LLM-MAS) have emerged as a powerful paradigm for tackling complex tasks by breaking them into collaborative workflows of specialized LLM-powered agents. However, deploying such multi-agent workloads at scale poses significant system challenges. Each user query spawns an iterative pipeline of LLM calls, greatly amplifying resource consumption compared to single-turn queries. In resource-constrained cloud settings, these workflows face non-deterministic and input-dependent costs at decode stage, heavy-tailed multi-model requirements with memory fragmentation and over-provisioning, and cross-cluster scheduling trade-offs. 
We present \projtitle, a workload-aware scheduling system designed for LLM-MAS serving under strict GPU budgets. \projtitle explicitly leverages agent semantics and roles: it predicts the output length and memory usage of each stage and uses this prediction to drive a hierarchical scheduler. At the node level, \projtitle enables dynamic multi-model co-location via hierarchical weight caching and elastic memory provisioning. At the cluster level, it performs latency-aware routing to avoid cold-start delays and memory overloads. At the global level, it enforces workflow-aware prioritization to minimize head-of-line blocking for interactive tasks. 
Across prototype experiments and trace-driven simulations, \projtitle
reduces KV-reservation HBM by 67.2\% and improves high-contention
SLO attainment over EDF by 23.6 percentage points.

\end{abstract}

\section{Introduction}
\label{sec:introduction}

Large Language Models (LLMs) are reshaping modern AI services~\cite{chen2023agentversefacilitatingmultiagentcollaboration,guo2025deepseek}. 
With recent open-source models (e.g., LLaMA~\cite{touvron2023llamaopenefficientfoundation}, DeepSeek~\cite{guo2025deepseek}, Qwen~\cite{yang2025qwen3}) demonstrating exceptional general-purpose reasoning and generation capabilities, \emph{LLM-based Multi-Agent Systems} (LLM-MAS) have emerged as a promising paradigm for tackling complex tasks~\cite{hong2024metagpt, wu2023autogenenablingnextgenllm}. 
In this architecture, a user request is decomposed into a structured workflow where specialized agents (e.g., planner, solver, critic) collaborate via natural language messages. This collaborative approach integrates memory, planning, and communication modules, enabling enhanced autonomy and adaptability that often outperform monolithic models in both robustness and quality~\cite{du2024improving, guoLargeLanguageModel2024,wang2024mixture}.

Running LLM-MAS at scale introduces fundamental challenges for inference serving infrastructure, particularly in private or on-premises deployments that lack the near-infinite elasticity of public clouds. 
In such resource-constrained settings~\cite{sheng2023flexgenhighthroughputgenerativeinference, theaibrixteam2025aibrixscalablecosteffectivelarge}, GPU capacity is typically constrained, fragmented, and distributed.
These constraints impose a critical trade-off between Quality-of-Service (QoS) and resource efficiency--a conflict significantly exacerbated by the \textit{dependency-coupled} nature of multi-agent workflows. 
Consider a sequential workflow (e.g., Planner $\to$ Coder $\to$ Reviewer): keeping paused agents' context resident in GPU memory ensures low latency but blocks other requests, while reclaiming it to free capacity incurs reloading overheads that violate SLOs. 
Consequently, naive scheduling~\cite{wu2024fastdistributedinferenceserving} or static provisioning--effective for independent queries--fails to resolve this magnified contention in LLM-MAS. 

\textbf{Challenge 1: Non-deterministic and input-dependent execution cost}.
LLM inference is autoregressive and produces outputs of variable and hard-predictable length.
In a multi-agent workflow, the latency and memory footprint of each stage depend on its output, which is unknown in advance. Without accurate cost estimates, the scheduler cannot properly prioritize requests or allocate memory, leading to head-of-line (HoL) blocking where long reasoning steps delay short ones. This phenomenon is further exacerbated in LLM-MAS, where decoding costs vary with factors such as agent's role, type of service/tool, and reasoning strategy (e.g., Chain-of-Thought).

\textbf{Challenge 2: Long-tail usage patterns and memory contention}.
In multi-agent pipelines, model invocation typically follows a heavy-tailed pattern: only a few models are invoked frequently, whereas most are used rarely.
Assigning a dedicated GPU to each model results in resource underutilization, while loading models on demand incurs cold-start costs (e.g., weight loading) that breach interactive latency SLOs.
Co-locating multiple models on a single GPU mitigates underutilization but intensifies memory contention.
Beyond static model parameters, the KV cache size grows with output length, creating dynamic pressure on limited GPU memory. 

\textbf{Challenge 3: Cross-cluster routing trade-offs (latency vs. resource readiness)}. 
In distributed settings, GPU clusters differ in both network delay and resource availability. 
Routing to the topologically nearest cluster minimizes network latency, but that cluster may lack the cached model weights or sufficient GPU memory to serve the request immediately.
A more distant cluster may demonstrate high \textit{resource readiness} (i.e., the model is loaded and memory is available), but the higher round-trip time increases end-to-end latency, especially time-to-first-token.
Thus, inefficient routing can propagate delays throughout the pipeline, slowing all subsequent stages.

Existing solutions only partially address these challenges~\cite{duanMuxServeFlexibleSpatialtemporal2024, patke2024queue, wu2024fastdistributedinferenceserving}: prediction-aware schedulers (\textsc{FastServe}~\cite{wu2024fastdistributedinferenceserving}) and multi-model systems (MuxServe~\cite{duanMuxServeFlexibleSpatialtemporal2024}, QLM~\cite{patke2024queue}) target single-turn queries with fixed partitioning; cross-cluster frameworks (SkyServe~\cite{maoSkyServeServingAI2025}, AIBrix~\cite{theaibrixteam2025aibrixscalablecosteffectivelarge}) follow static routing oblivious to model readiness or KV affinity. No existing system couples low-overhead stage-level cost estimation with cross-cluster allocation for agentic workflows.

To bridge this gap, we propose \projtitle, a workload-aware cross-cluster scheduling system for LLM-MAS. Unlike existing serving stacks that see only opaque LLM requests, \projtitle characterizes the workload along five agent-level dimensions that directly shape resource behavior: the agent's \textit{role and workflow position}, its \textit{tool-invocation intent} for the current step, the \textit{predicted per-stage output length and KV footprint}, the job's \textit{remaining workflow time}, and the \textit{model readiness} of each candidate cluster. Driven by this agent-level view of the workload, \projtitle coordinates decisions across three tiers: at the \textit{node level}, memory-safe multi-model colocation via hierarchical weight residency (GPU-CPU-disk) and prediction-guided elastic KV allocation; at the \textit{cluster level}, fitness-based dispatch that balances network latency, model readiness, and KV feasibility; and at the \textit{global level}, workflow-aware preemptive Shortest-Remaining-Time-First queueing that protects latency-critical stages without starving long analytic jobs.

This paper makes the following contributions.

\textbf{i) Agent-aware cost prediction.}
\projtitle presents a semantic- and structure-aware stage cost predictor that distinguishes concise tool-invocation outputs from extended Chain-of-Thought generations, lowering token-length prediction error by about 19\% and yielding accurate per-stage runtime and tool-usage estimates for scheduling.

\textbf{ii) Multi-agent spatio-temporal multiplexing.}
\projtitle implements a node-level runtime that co-optimizes hierarchical model placement across GPU/CPU/disk and prediction-driven elastic KV cache allocation, allowing memory-efficient colocation of many specialized models under strict GPU memory.

\textbf{iii) Feasibility-aware cross-cluster scheduling.}
\projtitle unifies network latency, per-cluster model readiness, and predicted KV footprint into a single fitness score that drives cross-cluster dispatch, coupled with boundary-preemptive Shortest-Remaining-Time-First to curb dependency-induced blocking under mixed SLOs.

\textbf{iv) Implementation and evaluation.}
Through a complete system prototype on a physical GPU cluster and trace-driven simulations,
\projtitle reduces prediction MAE by 19.2\%,
reduces KV-reservation HBM by 67.2\%, and improves high-contention
SLO attainment by 23.6 percentage points.

\section{Background \& Motivation}
\label{sec:background}

\noindent\textbf{Multi-agent serving workflows under mixed SLOs}.
LLM-MAS processes each user request as a structured workflow of dependent stages {\inlinemathsize $\{T_1,\dots, T_n\}$}. 
Each stage {\inlinemathsize $T$} represents a logical step executed by a specialized agent using a specific model {\inlinemathsize $M(T)$}, which may involve natural language generation or external tool invocation~\cite{chen2023agentversefacilitatingmultiagentcollaboration, hong2024metagpt, wu2023autogenenablingnextgenllm, wang2024mixture}.
Dependency coupling means upstream delays propagate to block subsequent steps, while resource costs vary systematically by agent role--e.g., concise tool calls versus extended Chain-of-Thought reasoning~\cite{yao2022react}.

In LLM-MAS with SLOs, \textit{user-facing stages} prioritize Time-to-First-Token (TTFT) for interactivity, whereas \textit{internal reasoning steps} demand low Time-to-Last-Token (TTLT) to rapidly unblock downstream dependencies.
Scheduling is distinctively challenging: dependency-induced head-of-line blocking~\cite{patke2024queue} means slow internal stages delay the final response, and cross-cluster routing must balance network latency against resource readiness~\cite{maoSkyServeServingAI2025,theaibrixteam2025aibrixscalablecosteffectivelarge}.
Unlike stateless microservices~\cite{gujarati2020clockwork}, LLM-MAS requires jointly managing these heterogeneous urgencies and heavy, stateful memory footprints (KV cache) under strict constraints.

\cpara{LLM inference cost model}.
An LLM (especially a decoder-only model) processes each request in two phases: a \textit{prefill} (or prompt processing) phase and an autoregressive \textit{decode} phase. Let {\inlinemathsize $P$} be the prompt length (i.e., number of input tokens) and {\inlinemathsize $L$} be the output length (number of generated tokens). The time for the prefill {\inlinemathsize $T_{\text{prefill}}(P)$} increases roughly with {\inlinemathsize $P$}, and the time for decoding rises with {\inlinemathsize $L$}. For large {\inlinemathsize $L$}, the decode time mainly dominates the total latency. The TTLT can be described as {\inlinemathsize $ T_{\mathrm{TTLT}} = T_{\mathrm{prefill}}(P) + L \cdot t_{\mathrm{decode}}$},
where {\inlinemathsize $t_{\text{decode}}$} is the average per-token generation time for the model on given hardware. This means that if we underestimate how long {\inlinemathsize $L$} will be, we risk underestimating the runtime of the job.

\cpara{KV cache memory pressure}.
At inference, memory is consumed not only by model weights but also by the KV cache, which stores hidden states for all processed tokens. 
KV cache usage scales linearly with the total sequence length {\inlinemathsize $(P + L)$}, amplified by model architectural factors (e.g., layers, heads) and precision.
This couples runtime and memory costs: generating a long output occupies GPU compute longer while consuming proportionally more memory, and under high concurrency the aggregate KV footprint can quickly exceed physical GPU limits.

\begin{figure}[t!]
  \centering
  \begin{minipage}[t]{0.48\linewidth}
    \centering
    \includegraphics[width=\linewidth]{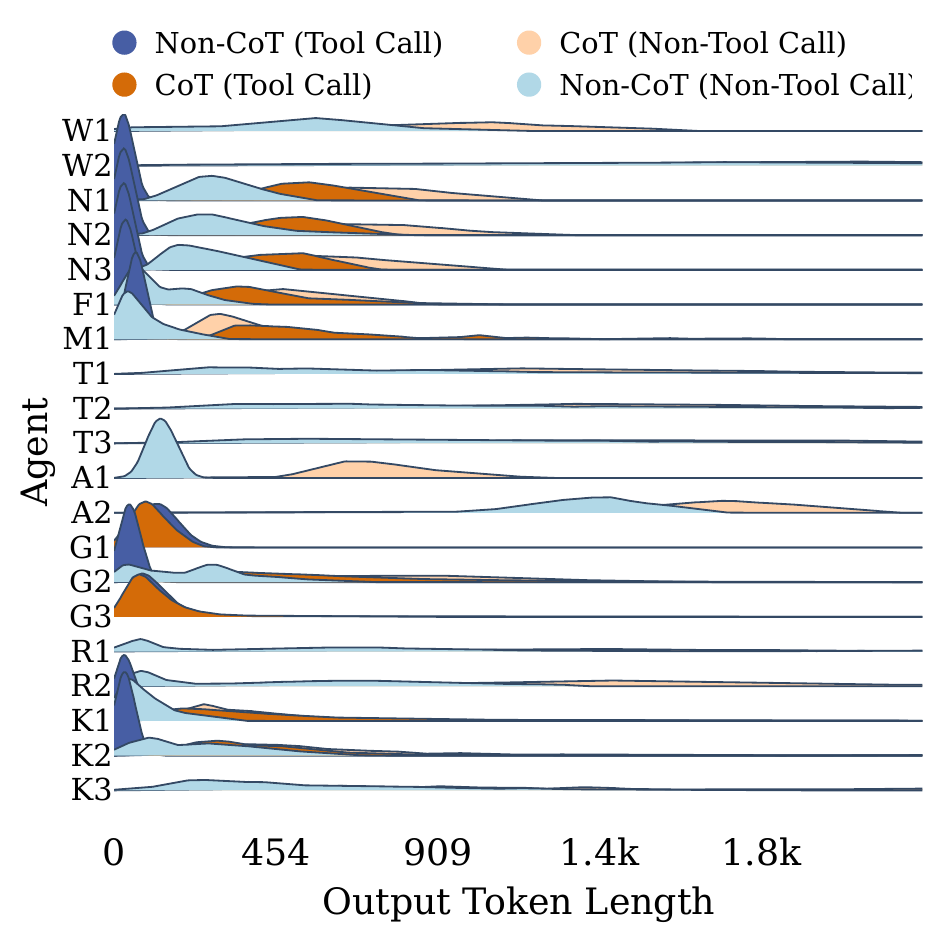}
    \caption{Output-token length distributions under non-CoT and CoT settings, tool-call and non-tool-call.}
    \label{fig:completion_tokens_length_distribution}
  \end{minipage}
  \hfill 
  \begin{minipage}[t]{0.48\linewidth}
    \centering
    \includegraphics[width=\linewidth]{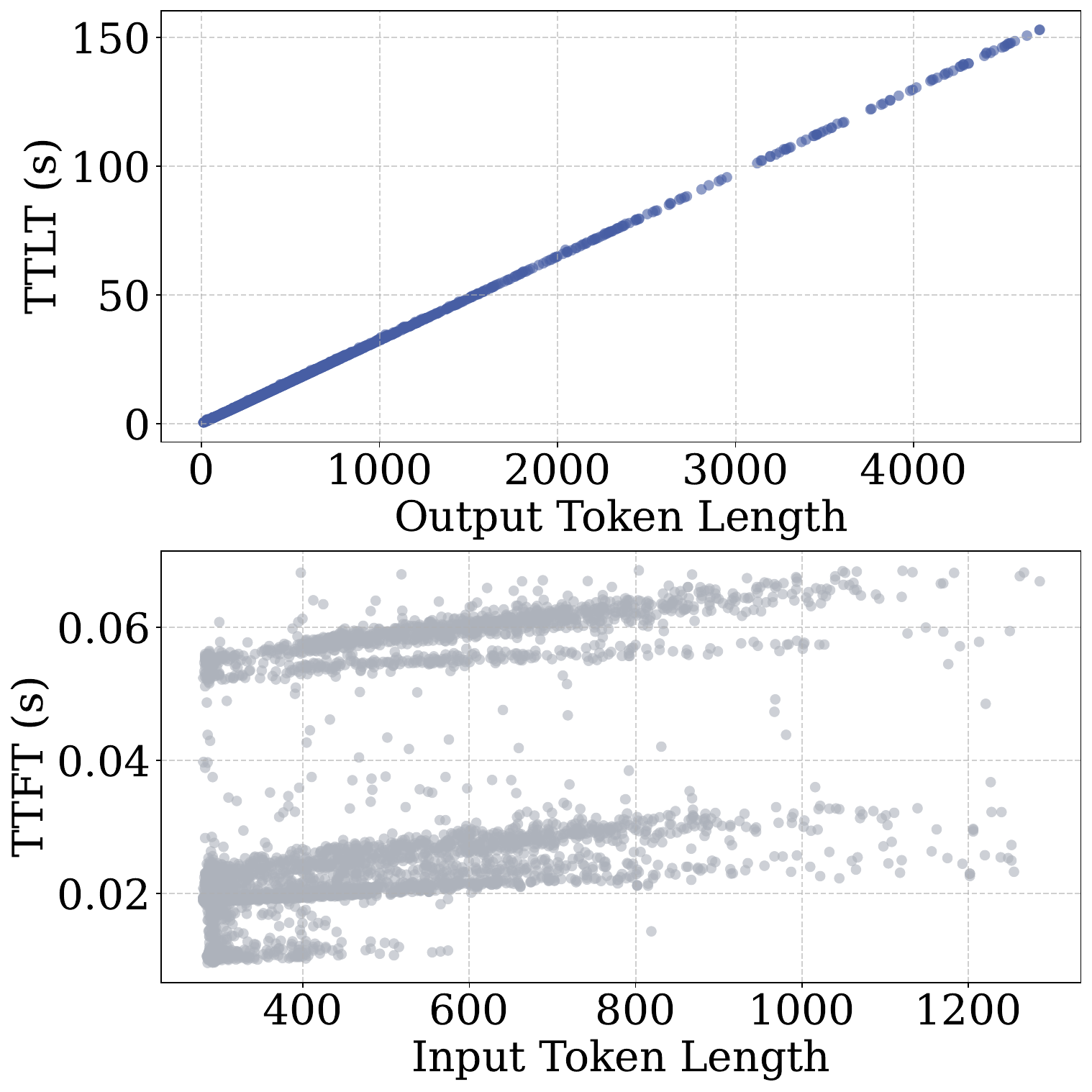}
    \caption{Relationships between output token length and TTLT (latency), and input token length and TTFT.}
    \label{fig:llm_ma_delay}
  \end{minipage}
  \vspace{-4pt}
\end{figure}

State-of-the-art serving engines~\cite{kwonEfficientMemoryManagement2023, yu2022orca, zhengSGLangEfficientExecution2024} employ paging techniques to optimize memory layout.
While PagedAttention~\cite{kwonEfficientMemoryManagement2023} effectively eliminates internal fragmentation via non-contiguous allocation, it does not resolve the fundamental \textit{capacity constraint}.
Under heavy load with long-context agents, the aggregate KV footprint can still exceed physical GPU limits.
Consequently, the system is forced to perform strict admission control, blocking new requests or preempting active ones, rather than relying on low-level memory compaction.

\cpara{Multi-agent deployments exacerbate memory challenges}. 
LLM-MAS typically utilizes specialized model variants for different agent roles. 
To avoid cold-start latency, serving systems often maintain multiple models in a \textit{resident state} (i.e., weights loaded in GPU or host memory).
While this ensures model readiness, it incurs strict resource costs: each resident model occupies memory not only for its parameters but also for associated runtime overheads (e.g., CUDA graphs, context buffers).
This static footprint competes directly with the dynamic capacity required for the KV cache. 


\cpara{Motivation.} To concretize these challenges, we performed detailed measurements on representative multi-agent workloads. 

\textbf{Observation-1}: \textit{Agent roles, tool invocation, and reasoning mode strongly determine output length and latency.}
Output token distributions vary widely across workflow stages, shown in Figure~\ref{fig:completion_tokens_length_distribution}: tool-oriented agents typically emit short, structured outputs, while user-facing or reasoning-heavy agents generate long free-form text. Enabling Chain-of-Thought reasoning further shifts output toward heavy-tailed distributions. 
We observe that output length strongly correlates with TTLT (inference latency), while prompt length mainly affects TTFT, shown in Figure~\ref{fig:llm_ma_delay}. 
Consequently, accurate per-stage output-length prediction is essential to anticipate long-running stages, mitigate head-of-line blocking, and perform memory-safe admission, since KV cache usage scales with sequence length.

\begin{figure}[t]
  \centering
  \includegraphics[width=0.3\linewidth, align=c]{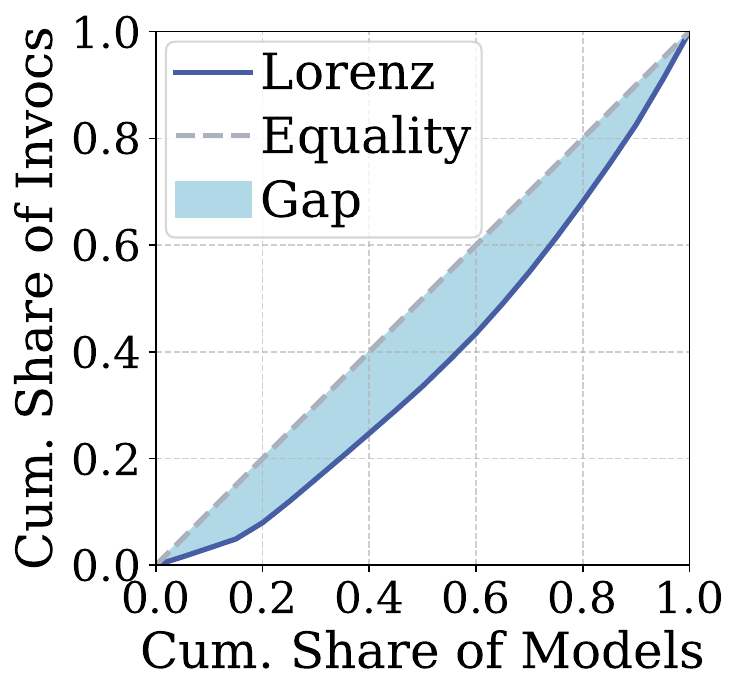}
  \hfill
  \includegraphics[width=0.68\linewidth, align=c]{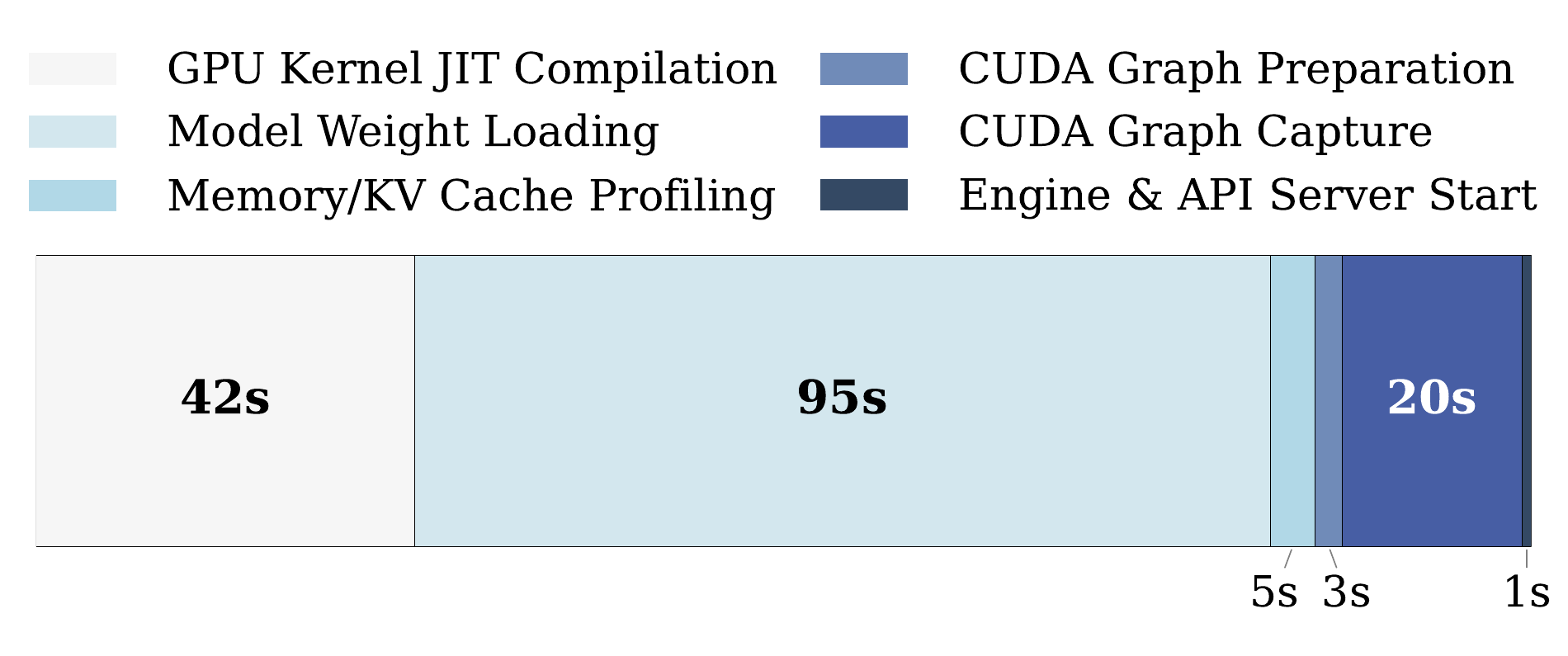}
  \caption{Model-invocation long tail (left) from Chatbot Arena  traces~\cite{zhengJudgingLLMasajudgeMTbench2023}. Cold-start breakdown (right) for an 8B model under standard serving.}
  \label{fig:multi_model_motivation}
\end{figure}

\textbf{Observation-2}: \textit{Multi-model demand is highly skewed, and cold starts fundamentally conflict with latency SLOs.}
Workload traces exhibit a pronounced long tail, shown in Figure~\ref{fig:multi_model_motivation}. Many specialized models are invoked infrequently but remain necessary. Static per-model GPU allocation wastes capacity, while on-demand loading incurs cold-start delays of tens of seconds, far exceeding interactive SLOs ~\cite{zhengJudgingLLMasajudgeMTbench2023}. Naive multi-model colocation further exacerbates GPU memory contention and fragmentation, motivating predictive, hierarchical model residency and elastic memory management.


\begin{figure}[t]
  \centering
  \includegraphics[width=\linewidth]{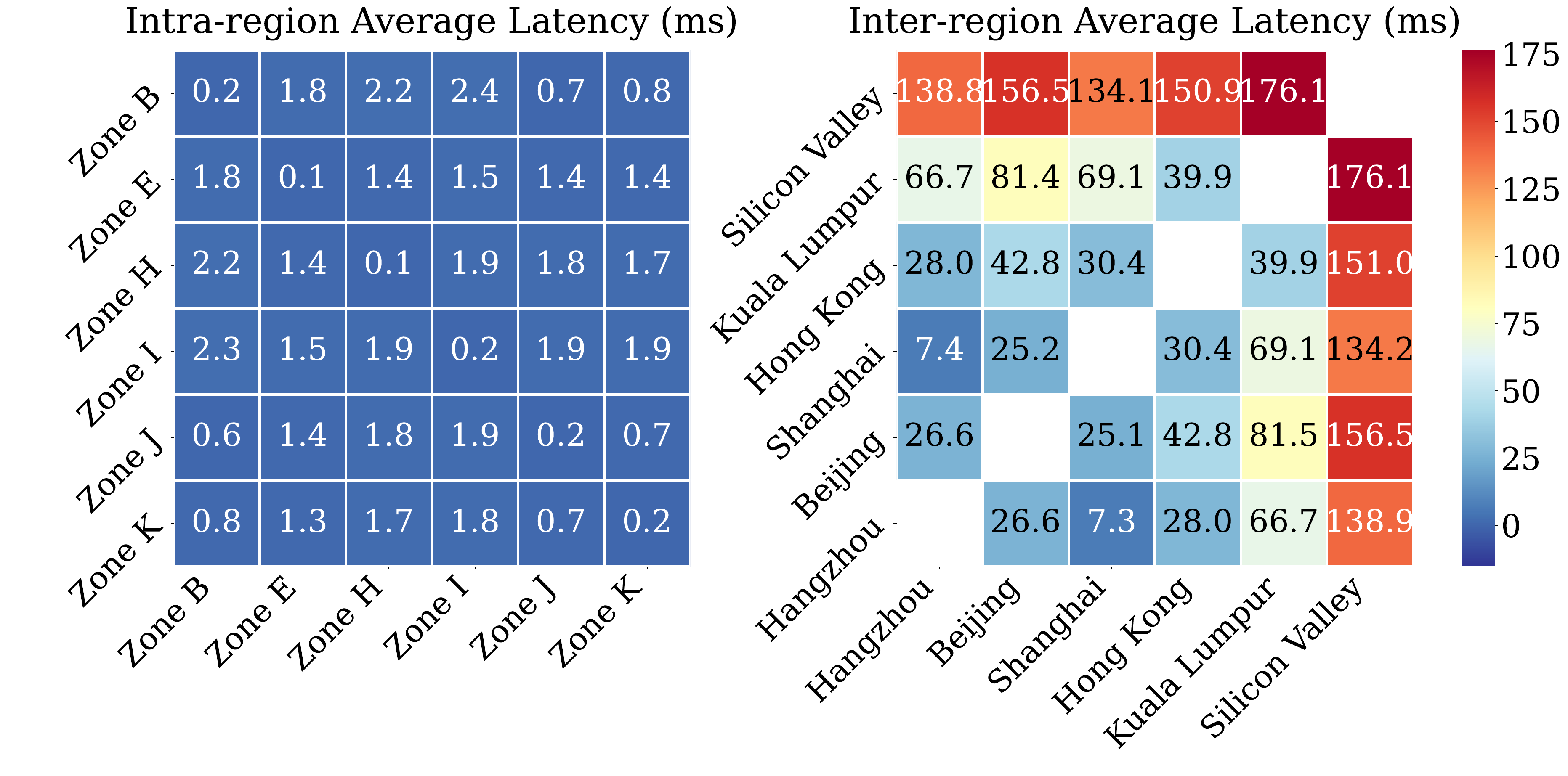}
    \vspace{-0.4em}
    \caption{Intra- and inter-region network latency reported by~\cite{alibabacloud_network_perf_2025}.}
  \label{fig:aliyun_latency}
  \vspace{-4pt}
\end{figure}

\textbf{Observation-3:} \textit{Cross-cluster routing must balance network latency with resource readiness.}
Although routing to nearby clusters minimizes network round-trip time (RTT), Figure~\ref{fig:aliyun_latency} implies that local clusters may lack the required model weights or sufficient KV capacity.
In such scenarios, dispatching execution to a remote cluster with a \textit{ready} model yields lower end-to-end latency, provided that the inter-region RTT (typically tens to hundreds of milliseconds) is orders of magnitude smaller than the local cold-start overhead (tens of seconds).
Consequently, effective scheduling must jointly optimize for network proximity and \textit{resource readiness}--specifically, both model residency and KV availability.

In summary, these observations indicate that high-performance LLM-MAS serving hinges on (i) precise workload forecasting at each stage, (ii) coordinated multi-model GPU memory orchestration, and (iii) intelligent, feasibility-aware cluster-wide scheduling. \projtitle meets these needs through a unified and prediction-guided scheduling pipeline.
\section{System Design}
\label{sec:design}

\projtitle enables LLM-MAS execution in multi-cluster environments by jointly deciding \textit{when} and \textit{where} to place each workflow stage {\inlinemathsize $T$}, selecting cluster and GPU while managing model residency and GPU memory. 

\begin{figure*}[t!]
  \centering
  \includegraphics[width=1\textwidth]{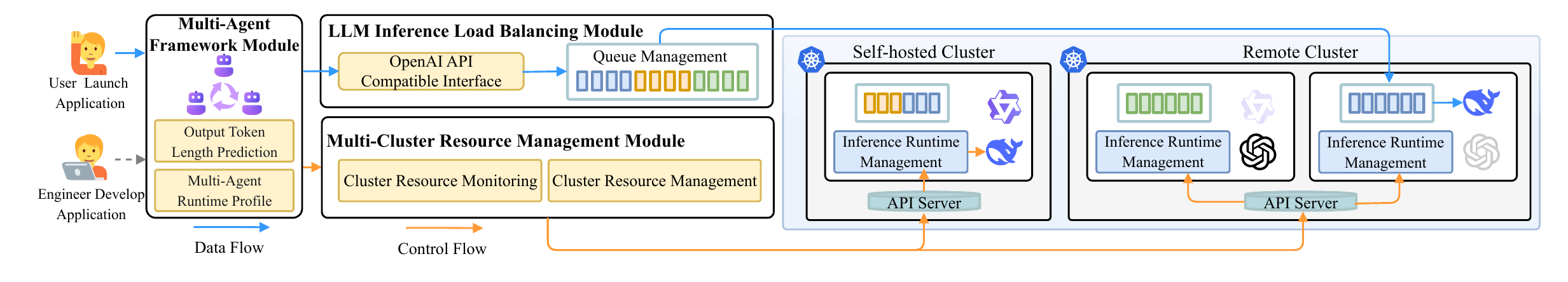}
  \vspace{-0.4em}
  \caption{The system architecture and workflow of \projtitle.}
  \label{fig:architecture}
  \vspace{-1em}
\end{figure*}

\subsection{Overview}
\label{sec:design_overview}

\projtitle executes LLM-MAS through a stage-driven closed-loop control pipeline, as illustrated in Figure~\ref{fig:architecture}. Each workflow stage {\inlinemathsize $T$} is an independent scheduling unit that passes through five phases:

\cpara{Agent-context observation.}  
When a stage is created, \projtitle intercepts the invocation and extracts a compact descriptor capturing the agent role, workflow position, tool availability, and a \textit{semantic embedding} of the input context. In multi-cluster deployments, it also records the source cluster and any policy constraints that restrict placement.

\cpara{Cost prediction.} 
A lightweight predictor at the \textit{dispatch gateway} estimates the expected output length {\inlinemathsize $\hat{L}(T)$}, the KV cache requirement {\inlinemathsize $\hat{R}_{\mathrm{kv}}(T)$}, and the tool-use probability {\inlinemathsize $\hat{p}_{\mathrm{tool}}(T)$}, incurring negligible latency relative to network RTT.

\cpara{Scheduling decision.}
The global scheduler filters infeasible nodes by policy and memory constraints, then dispatches the stage to the node that maximizes a fitness score balancing network latency, time-to-start, and KV cache feasibility. Stages are ordered by workflow-aware Shortest-Remaining-Time-First with preemption for latency-critical work.

\cpara{Node-level execution and memory management.} 
The runtime coordinates model loading, KV cache allocation with explicit memory accounting, and minimum-disruption reclamation under memory pressure, reporting disruption costs back to the scheduler. 

\cpara{Post-execution profiling.} 
Actual output length, latency, and memory usage are recorded to incrementally calibrate the predictor and re-prioritize remaining stages.

\subsection{Agent-Aware Cost Prediction}
\label{sec:design_prediction}

\begin{figure}[t!]
  \centering
  \includegraphics[width=0.5\textwidth]{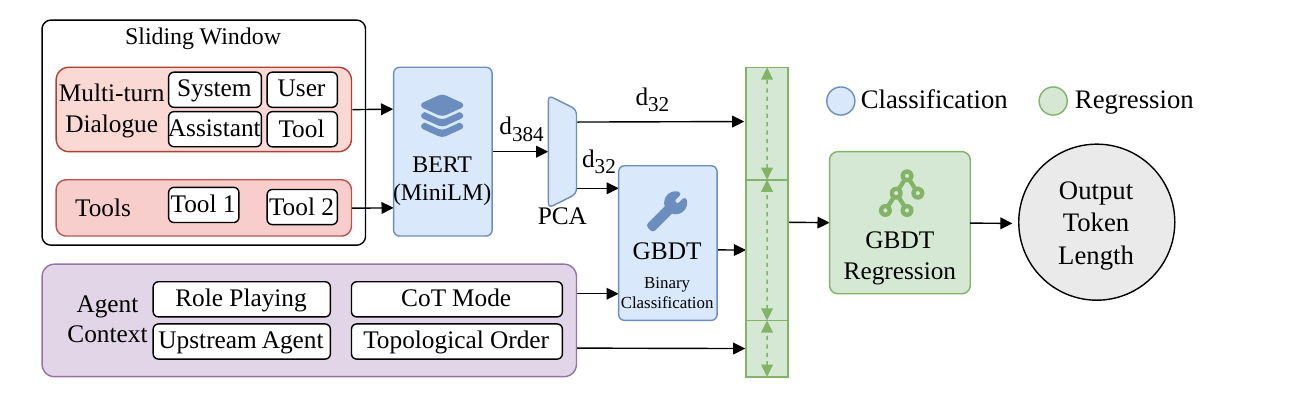}
  \caption{Agent-aware output-length prediction architecture in \projtitle.}
  \label{fig:agent_output_pred_arch}
  \vspace{-1em}
\end{figure}


\cpara{Feature representation.}
Input features {\inlinemathsize $\boldsymbol{x}(T)$} concatenate \textit{structured features} (including agent role, workflow position, invocation index, and tool availability) and \textit{semantic features} derived from the input text.
To handle long prompts efficiently without the overhead of full-scale attention, \projtitle applies a sliding-window encoding using a lightweight BERT model (MiniLM) and aggregates embeddings via mean pooling.
This yields a compact semantic representation {\inlinemathsize $\boldsymbol{x}_{\mathrm{sem}}(T)$} that captures context complexity.

\cpara{Two-phase prediction pipeline.}
We observe that tool-invoking phases typically produce short structured outputs, while user-facing stages generate longer free-form text, resulting in a bimodal output-length distribution. This motivates a two-stage predictor (Figure~\ref{fig:agent_output_pred_arch}) that explicitly models tool-invoking intent.

\projtitle first employs a lightweight classifier to estimate the probability that stage {\inlinemathsize $T$} triggers a tool call:
\begin{equation}
    \hat{p}_{\mathrm{tool}}(T)=\Pr\!\left[\mathrm{tool}(T)=1 \mid \boldsymbol{x}(T)\right]
\end{equation}
To ensure the predicted confidence aligns with empirical frequencies, we apply isotonic regression to calibrate the classifier outputs, providing a reliable continuous signal to guide downstream estimation. If no tools are available, we set {\inlinemathsize $\hat{p}_{\mathrm{tool}}(T)=0$}.

\projtitle then predicts the output length {\inlinemathsize $\hat{L}(T)$} using structured features {\inlinemathsize $\boldsymbol{x}_{\mathrm{str}}(T)$}, semantic features {\inlinemathsize $\boldsymbol{x}_{\mathrm{sem}}(T)$}, and the predicted tool-invoking probability {\inlinemathsize $\hat{p}_{\mathrm{tool}}(T)$}.
To capture role-specific generation patterns, we train per-role regressors when training data are available; otherwise, we fall back to a shared global model. To mitigate the impact of heavy-tailed output distributions, we perform a regression on {\inlinemathsize $\log(1+L(T))$} (where {\inlinemathsize $L(T)$} is the length of the ground-truth output) 
and apply the inverse transform at inference time.

\cpara{Translation to system metrics.}
\projtitle translates the predicted length into execution-time and memory estimates using calibrated per-model microbenchmarks. Let {\inlinemathsize $t_{\mathrm{pre}}(P,M)$} be the prefill latency for prompt length {\inlinemathsize $P$} and {\inlinemathsize $t_{\mathrm{dec}}(M)$} be the average per-token decode latency. The estimated execution time is:
\begin{equation}
    \widehat{T}_{\mathrm{exec}}(T) =
    t_{\mathrm{pre}}(P(T), M(T)) + t_{\mathrm{dec}}(M(T))\cdot \hat{L}(T)
    \label{eq:design_exec_time}
\end{equation}
which captures the inference cost. Tool execution latency is profiled as an external workflow stage and added to the remaining-time profile when traces expose it; the current runtime, however, does not reserve CPU or network resources for external tools. The KV cache requirement is estimated as:
\begin{equation}
    \hat{R}_{\mathrm{kv}}(T) = \alpha(M(T))\cdot \bigl(P(T)+\hat{L}(T)\bigr)
    \label{eq:kv_demand}
\end{equation}
where {\inlinemathsize $\alpha(M)$} denotes the model-specific memory footprint per token. These estimates provide the scheduler with explicit signals for execution-time ordering and memory-feasible placement.

\RestyleAlgo{ruled}
\begin{algorithm}[t]
  \caption{Hierarchical Weight Residency and Eviction}
  \label{alg:hierarchical_cache_eviction}
  \footnotesize
  \KwIn{Model $m$, Capacities $C_{\mathrm{gpu}}, C_{\mathrm{cpu}}, C_{\mathrm{disk}}$}
  \KwOut{Model $m$ becomes GPU-ready}
  $S_m \leftarrow \mathrm{GetModelSize}(m)$\;
  $Loc \leftarrow \mathrm{LocateModel}(m)$\;
  
  \If{$Loc = \mathrm{GPU}$}{
      $\mathrm{LRU.Update}(\mathrm{GPU}, m)$\;
      \Return{$\mathrm{Success}$}\;
  }

  \While{$\mathrm{UsedGPU}() + S_m > C_{\mathrm{gpu}}$}{
      $v \leftarrow \mathrm{LRU.PeekLast}(\mathrm{GPU})$\;
      $\mathrm{OffloadToHost}(v)$\; 
      $\mathrm{LRU.Remove}(\mathrm{GPU}, v)$\;
      $\mathrm{LRU.Update}(\mathrm{CPU}, v)$\;
  }

  \If{$Loc \in \{\mathrm{Disk}, \mathrm{Remote}\}$}{
    \While{$\mathrm{UsedCPU}() + S_m > C_{\mathrm{cpu}}$}{
      $v \leftarrow \mathrm{LRU.PeekLast}(\mathrm{CPU})$\;
      $S_v \leftarrow \mathrm{GetModelSize}(v)$\;
      \If{$\mathrm{UsedDisk}() + S_v \le C_{\mathrm{disk}}$}{
        $\mathrm{MoveToDisk}(v)$\;
      }\Else{
        $\mathrm{DeleteFromDisk}(v)$\;
      }
      $\mathrm{LRU.Remove}(\mathrm{CPU}, v)$\;
    }
    $\mathrm{LoadToCPU}(m)$\;
  }

  $\mathrm{LoadToGPU}(m)$\;
  $\mathrm{LRU.Update}(\mathrm{GPU}, m)$\;
  \Return{$\mathrm{Success}$}\;
\end{algorithm}

\subsection{Node-Level Execution and Memory Management}
\label{sec:design_multiplex}
\projtitle employs a node-level runtime manager to enable \textit{memory-feasible multi-model colocation} on a single GPU. The design combines \textit{temporal multiplexing}, which amortizes model activation costs, with \textit{spatial multiplexing}, which increases KV cache concurrency under a shared GPU memory budget. Crucially, the runtime exports readiness and KV-feasibility signals that allow the cross-cluster scheduler to avoid routing latency-sensitive stages to nodes that would incur cold starts or disruptive memory admissions.

\cpara{Model readiness abstraction and memory accounting.}
\projtitle employs a \textit{model readiness abstraction} to capture the activation cost of a model on a node. 
A model may be in one of five states:  
\textit{Running} (weights and execution context resident on GPU), 
\textit{Sleeping} (weights offloaded to host memory, but lightweight GPU runtime contexts--e.g., CUDA graphs and JIT kernels--are retained to accelerate reloading),
\textit{CPU-resident} (weights cached in host memory without preserved GPU context), 
\textit{Disk-resident}, or \textit{Remote}. 
For each model, the runtime reports both its readiness state and an activation latency estimated via a \textit{profiled bandwidth model} (i.e., $T_{\mathrm{act}} \approx \text{Size} / \mathrm{BW}_{\mathrm{tier}}$), allowing switching costs to be compared uniformly across clusters.

Multi-model serving is constrained by a shared GPU memory budget in which persistent warm contexts directly compete with KV cache capacity. Rather than reserving a fixed fraction of memory for KV cache, \projtitle enforces \textit{explicit memory accounting} with admission-time feasibility checks.

Let {\inlinemathsize $M_{\mathrm{total}}$} denote the total GPU memory available to the runtime, {\inlinemathsize $M_{\mathrm{kv}}$} the current KV cache usage, {\inlinemathsize $\mathcal{S}$} the set of warm models (Running or Sleeping), {\inlinemathsize $M_{\mathrm{ctx}}^{k}$} the persistent context footprint of model {\inlinemathsize $k$}, and {\inlinemathsize $M_{\mathrm{other}}$} non-model overheads.
The reserved non-KV footprint is
{\inlinemathsize $M_{\mathrm{res}} \;=\; \sum_{k \in \mathcal{S}} M_{\mathrm{ctx}}^{k} + M_{\mathrm{other}}$}
and memory safety is enforced by {\inlinemathsize $M_{\mathrm{kv}} + M_{\mathrm{res}} \le M_{\mathrm{total}}$}.

At admission time, the runtime checks whether the incoming stage’s additional KV demand can be safely allocated. If feasible, the stage is admitted and the KV admission headroom {\inlinemathsize $R_{\mathrm{kv}}^{\mathrm{head}}(N)= M_{\mathrm{total}}-M_{\mathrm{res}}-M_{\mathrm{kv}}$}
is reported as a scheduling signal; otherwise, the runtime rejects the stage or triggers minimum-impact memory coordination, reporting infeasibility and the associated disruption cost.

\cpara{Temporal multiplexing via hierarchical weight residency.}
To amortize cold-start overheads, the runtime maintains a hierarchical weight residency path (GPU $\to$ Host RAM $\to$ Local Disk $\to$ Remote Storage), with LRU eviction policies applied at each tier. 
When a stage requests model {\inlinemathsize $m$}, the runtime activates it from the fastest available tier. 
If the GPU memory is full, the least recently used model is offloaded to host memory (transitioning to \textit{Sleeping} or \textit{CPU-resident} state).
If host memory is subsequently insufficient, cold weights are evicted to local disk (or discarded if disk capacity is exceeded) to make room.
Algorithm~\ref{alg:hierarchical_cache_eviction} summarizes this cascading load-and-evict process.

The runtime periodically reports the current warm-set composition and model activation-cost estimates. The global scheduler exploits these signals during cross-cluster routing, preferentially dispatching latency-sensitive stages to clusters where the required model is already warm.

\cpara{Spatial multiplexing with virtual-memory KV cache.}
Temporal multiplexing alone necessitates conservative KV admission, which can underutilize GPU memory at low load. In nodes supporting CUDA virtual memory management (VMM), \projtitle enables \textit{spatial multiplexing} by constructing a shared virtual KV cache pool and mapping KV pages on demand.
We integrate \textit{kvcached}~\cite{ovg-projectKvcached2025} to allocate and reclaim KV pages via CUDA VMM, reducing fragmentation and enabling elastic KV provisioning. To maintain reclaimability under multi-model colocation, long-lived prefix caching is disabled in this mode unless explicitly required.

Although the virtual KV address space may exceed physical memory, \projtitle preserves safety through admission control and on-demand mapping--if physical allocation fails, the runtime rejects the stage or triggers controlled degradation and reports infeasibility to the scheduler, preventing out-of-memory failures. The runtime advertises VMM support as a capability-conditioned signal, which the scheduler treats as a hard feasibility constraint for high-concurrency stages and a soft preference for latency-sensitive interactive stages.

\cpara{Minimum-impact memory coordination as a routing signal.}
When KV admission fails, the runtime executes Algorithm~\ref{alg:memory_satisfaction} to derive a \textit{minimum-impact degradation plan}.
We define five degradation levels with ascending disruption costs:
(1) transitioning \textit{Idle-Running} models to \textit{Sleeping};
(2) evicting \textit{Sleeping} models;
(3) stopping pending sleep transitions;
(4) swapping out KV for \textit{Active} models; and
(5) aborting \textit{Active} executions.

The algorithm adopts a greedy strategy, iterating through resident engines {\inlinemathsize $\mathcal{E}$} sorted by priority (Idle $\to$ Sleeping $\to$ Active) and accumulating freed memory via state-dependent actions until the requirement {\inlinemathsize $R$} is met.

\RestyleAlgo{ruled}
\begin{algorithm}[t]
  \caption{Minimum-Impact Memory Coordination}
  \label{alg:memory_satisfaction}
  \footnotesize
  \KwIn{Required memory $R$, Resident engines $\mathcal{E}$}
  \KwOut{Degradation plan $P$, Interrupt flag $I_{\mathrm{active}}$}
  $M_{\mathrm{freed}}\leftarrow 0; P \leftarrow [\,]; I_{\mathrm{active}}\leftarrow \mathrm{False}$\;
  
  $\mathcal{E}_{sorted} \leftarrow \mathrm{SortByPriority}(\mathcal{E})$\;
  
  \For{$e \in \mathcal{E}_{sorted}$}{
    \If{$M_{\mathrm{freed}} \ge R$}{
        \textbf{break}\;
    }
    
    $Action \leftarrow \mathrm{DetermineBestAction}(e, P)$\;
    
    \If{$Action \in \{\mathrm{Preempt}, \mathrm{Abort}\}$}{
        $I_{\mathrm{active}}\leftarrow \mathrm{True}$\;
    }
    
    $M_{\mathrm{freed}}\leftarrow M_{\mathrm{freed}} + \mathrm{EstimateFreedMemory}(e, Action)$\;
    $P.\mathrm{append}((e, Action))$\;
  }
  
  \If{$M_{\mathrm{freed}} < R$}{
      \Return{$\mathrm{Failure}$}\;
  }
  \Return{$P, I_{\mathrm{active}}$}\;
\end{algorithm}

The runtime calculates total disruption penalty $C_{\mathrm{deg}}(N,T)$ as the \textit{aggregate restoration latency} required to execute a plan:
\begin{equation}
  C_{\mathrm{deg}}(N,T) = \sum_{(e,a)\in P} c(e,a) + \mathbf{1}[I_{\mathrm{active}}]\cdot c_{\mathrm{int}}
  \label{eq:deg_cost}
\end{equation}
Here, $c(e,a)$ is derived from profiled storage bandwidth (for model reloading) or compute throughput (for KV regeneration), while $c_{\mathrm{int}}$ corresponds to the SLO violation threshold.

\subsection{Workload-Aware Cross-Cluster Scheduling}
\label{sec:design_scheduling}

\projtitle employs a cross-cluster scheduling control loop that jointly performs routing, admission, and queueing for multi-agent workflow stages. Each placement must satisfy policy constraints, KV cache feasibility, and latency objectives. The scheduler uses (i) a feasibility-aware, fitness-based routing policy to select \textit{where} a stage runs, and (ii) a profile-driven remaining-time queueing policy to decide \textit{when} stages are dispatched under workflow dependencies.

\cpara{Cross-cluster fitness.}
For an arriving stage {\inlinemathsize $T$}, \projtitle{} predicts the KV demand {\inlinemathsize $\hat{R}_{\mathrm{kv}}(T)$} and applies a safety margin
{\inlinemathsize $R_{\mathrm{need}}(T)=(1+\rho)\cdot \hat{R}_{\mathrm{kv}}(T)$}.
Rather than a fixed heuristic, {\inlinemathsize $\rho$} is set from recent prediction errors: we maintain an EWMA of the relative underestimation
{\inlinemathsize $e=\max(0, R_{\mathrm{kv}}/\hat{R}_{\mathrm{kv}}-1)$} and choose {\inlinemathsize $\rho$} as a high quantile (e.g., 90--95\%) of {\inlinemathsize $e$} within a sliding window; in practice it typically falls in {\inlinemathsize $[0.1,0.3]$}.
The scheduler first filters candidate nodes using routing policies and feasibility signals reported by the runtime manager; only nodes that can safely admit {\inlinemathsize $R_{\mathrm{need}}(T)$} under memory constraints are considered.

Among feasible candidates, \projtitle selects the node that minimizes the expected stage completion latency
{\inlinemathsize $\widehat{T}_{\mathrm{e2e}}(N,T)\approx \mathrm{RTT}(\mathrm{src}(T),N)+T_{\mathrm{ready}}(N,T)+\widehat{T}_{\mathrm{exec}}(T)+\eta\,C_{\mathrm{deg}}(N,T)$}.
Since {\inlinemathsize $\widehat{T}_{\mathrm{exec}}(T)$} is node-invariant for a fixed model-engine pair, it does not affect the routing argmin and is used in the queueing policy below.
We therefore rank nodes by the equivalent compact score:
\begin{equation}
    S(N,T) = A(N,T) - \lambda\,T_{\mathrm{ready}}(N,T) - \mu\,C_{\mathrm{deg}}(N,T)
    \label{eq:fitness}
\end{equation}
where {\inlinemathsize $T_{\mathrm{ready}}(N,T)$} is the expected time-to-start on node {\inlinemathsize $N$} and {\inlinemathsize $C_{\mathrm{deg}}(N,T)$} is the disruption penalty reported by the runtime (Eq.~\ref{eq:deg_cost}).
The affinity term {\inlinemathsize $A(N,T)$} combines (i) network proximity using a decreasing transform of the measured RTT and (ii) KV cache fit via best-fit packing based on the runtime-reported headroom {\inlinemathsize $R_{\mathrm{kv}}^{\mathrm{head}}(N)$} (larger stages prefer nodes whose headroom is close to {\inlinemathsize $R_{\mathrm{need}}(T)$} among feasible candidates); for latency-sensitive interactive stages, we increase the weight on the network component.
Throughout, we apply robust min--max normalization with per-metric 5/95-percentile bounds over a recent window (clipped outside) to prevent outliers from dominating the score, and we report {\inlinemathsize $T_{\mathrm{ready}}$} and {\inlinemathsize $C_{\mathrm{deg}}$} in milliseconds (default {\inlinemathsize $\lambda{=}\mu{=}1$}).

The ready time term decomposes into queueing and activation latency:
\begin{equation}
    T_{\mathrm{ready}}(N,T)=T_{\mathrm{q}}(N,T)+T_{\mathrm{act}}(N,M(T))
    \label{eq:ready_time}
\end{equation}
where {\inlinemathsize $T_{\mathrm{q}}(N,T)$} is the estimated node-local queueing delay (smoothed with EWMA), and {\inlinemathsize $T_{\mathrm{act}}(N,M(T))$} is the model activation latency determined by the model readiness state on node {\inlinemathsize $N$}.

\RestyleAlgo{ruled}
\begin{algorithm}[t]
  \caption{Cross-Cluster Scheduling}
  \label{alg:unified_dispatch}
  \KwIn{Stage $T$, candidate nodes $\mathcal{N}$ (across clusters)}
  \KwOut{Selected node $N^{*}$}
  $(\hat{L},\hat{R}_{\mathrm{kv}})\leftarrow \mathrm{Predict}(T)$\;
  $R_{\mathrm{need}}\leftarrow (1+\rho)\cdot \hat{R}_{\mathrm{kv}}$\;
  $\mathcal{N}'\leftarrow \mathrm{FilterByPolicyAndFeasibility}(\mathcal{N},T,R_{\mathrm{need}})$\;
  $(\lambda,\mu)\leftarrow \mathrm{GetWeights}(T.\mathrm{class})$\;
  $N^{*}\leftarrow \varnothing; S^{*}\leftarrow -\infty$\;
  \For{$N\in \mathcal{N}'$}{
    $A \leftarrow \mathrm{Affinity}(\mathrm{RTT}(\mathrm{src}(T),N), R_{\mathrm{kv}}^{\mathrm{head}}(N), R_{\mathrm{need}}, T.\mathrm{class})$\;
    $(T_{\mathrm{ready}}, C_{\mathrm{deg}})\leftarrow \mathrm{RuntimeEstimate}(N,T)$\;
    $S \leftarrow A - \lambda\cdot T_{\mathrm{ready}} - \mu\cdot C_{\mathrm{deg}}$\;
    \If{$S>S^{*}$}{ $S^{*}\leftarrow S$; $N^{*}\leftarrow N$ }
  }
  \Return{$N^{*}$}\;
\end{algorithm}

\cpara{Profile-driven remaining-time queueing.}
\projtitle employs a two-level queueing architecture: a global queue that determines the dispatch order between jobs and node-local queues that schedule execution after placement. To mitigate dependency-induced head-of-line blocking, the global queue orders jobs by an estimate of their remaining workflow execution time.
For a job {\inlinemathsize $J$} currently executing stage {\inlinemathsize $T_k$}, the remaining time is estimated as
\begin{equation}
    \widehat{T}_{\mathrm{rem}}(J,k)=\widehat{T}_{\mathrm{exec}}(T_k)+\widehat{T}_{\mathrm{future}}(J,k)
    \label{eq:remaining_time}
\end{equation}
where {\inlinemathsize $\widehat{T}_{\mathrm{exec}}(T_k)$} is derived from the predicted output length {\inlinemathsize $\hat{L}(T_k)$} by Eq.~\ref{eq:design_exec_time}.
To account for branching and iterative behavior, \projtitle maintains a rolling execution profile for each workflow template and estimates the future term using a conditional median over recent executions:
\begin{equation}
    \widehat{T}_{\mathrm{future}}(J,k)\approx T^{(0.50)}_{\mathrm{next}}(\mathrm{state}(J,k))
    \label{eq:future_time}
\end{equation}
where {\inlinemathsize$\mathrm{state}(J,k)$} summarizes the workflow position using the agent role, stage template, invocation index, and a discretized tool-invoking intent score.

\cpara{Preemption and mixed SLOs.}
Under contention, \projtitle preempts background stages only at stage boundaries (between LLM invocations) and re-queues them, avoiding disruption of in-flight decoding. This choice is intentionally conservative: token- or iteration-level preemption can further reduce blocking for exceptionally long decoding phases, but requires tighter integration with the decoding engine and KV migration path. In our design, the common preemption path only updates scheduler metadata and preserves KV allocations; when memory pressure prevents preservation, the recovery latency is charged through the degradation term $C_{\mathrm{deg}}$ in Eq.~\ref{eq:deg_cost}.
To prevent oscillation due to estimation noise (e.g., in {\inlinemathsize $T_{\mathrm{ready}}$}), we apply hysteresis: preemption is triggered only when the predicted latency gain exceeds a threshold and a per-job cooldown expires; queueing-delay estimates are smoothed with EWMA.
KV cache allocations are preserved when feasible and otherwise reclaimed by the runtime using minimum-impact coordination, while aging gradually increases the effective priority of long-waiting background jobs to prevent starvation.

\section{Evaluation}
\label{sec:evaluation}

We evaluate \projtitle using both a physical GPU cluster and trace-driven simulations. We organize the results into three layers: (i) \textit{overall performance}, assessing end-to-end latency and SLO attainment under mixed workloads; (ii) \textit{resource efficiency}, quantifying GPU cost savings and memory overcommitment enabled by multi-model colocation; and (iii) \textit{component analysis}, evaluating the accuracy and runtime overhead of the agent-aware prediction module.

\subsection{Experimental Setup}
\label{sec:eval_setup}

\noindent\textbf{Testbed.}
We deploy \projtitle in a multi-cluster environment.
Our physical testbed comprises 32 GPU servers (each with 2 NVIDIA A100 GPUs, 128 CPU cores, 1TB RAM, and 2TB NVMe SSD local storage).
The software stack is Ubuntu 22.04.5 with Kubernetes (v1.31.2), NVIDIA Driver 580.105.08, and CUDA 13.0.
We use vLLM (v0.11.0) as the inference backend and extend its PagedAttention KV management with our elastic paging mechanism.

\cpara{Workloads.}
We replay a multi-agent workflow trace collected from 9 representative LLM-MAS applications (4 interactive and 5 batch-style), covering serial, parallel, loop, and supervisor-worker patterns.
The trace contains 46,769 jobs and 144,524 LLM-invocation stages (Table~\ref{tab:workload_trace}). Both synthetic and public-dataset workloads are generated from fixed application templates encoding common agent topologies (e.g., serial tool-use loops, supervisor-worker fan-out/fan-in, multi-step reasoning with refinement); they differ only in input sources, ensuring that train and test jobs share application logic but use distinct prompts.

\begin{figure}[t]
  \centering
  \includegraphics[width=\linewidth]{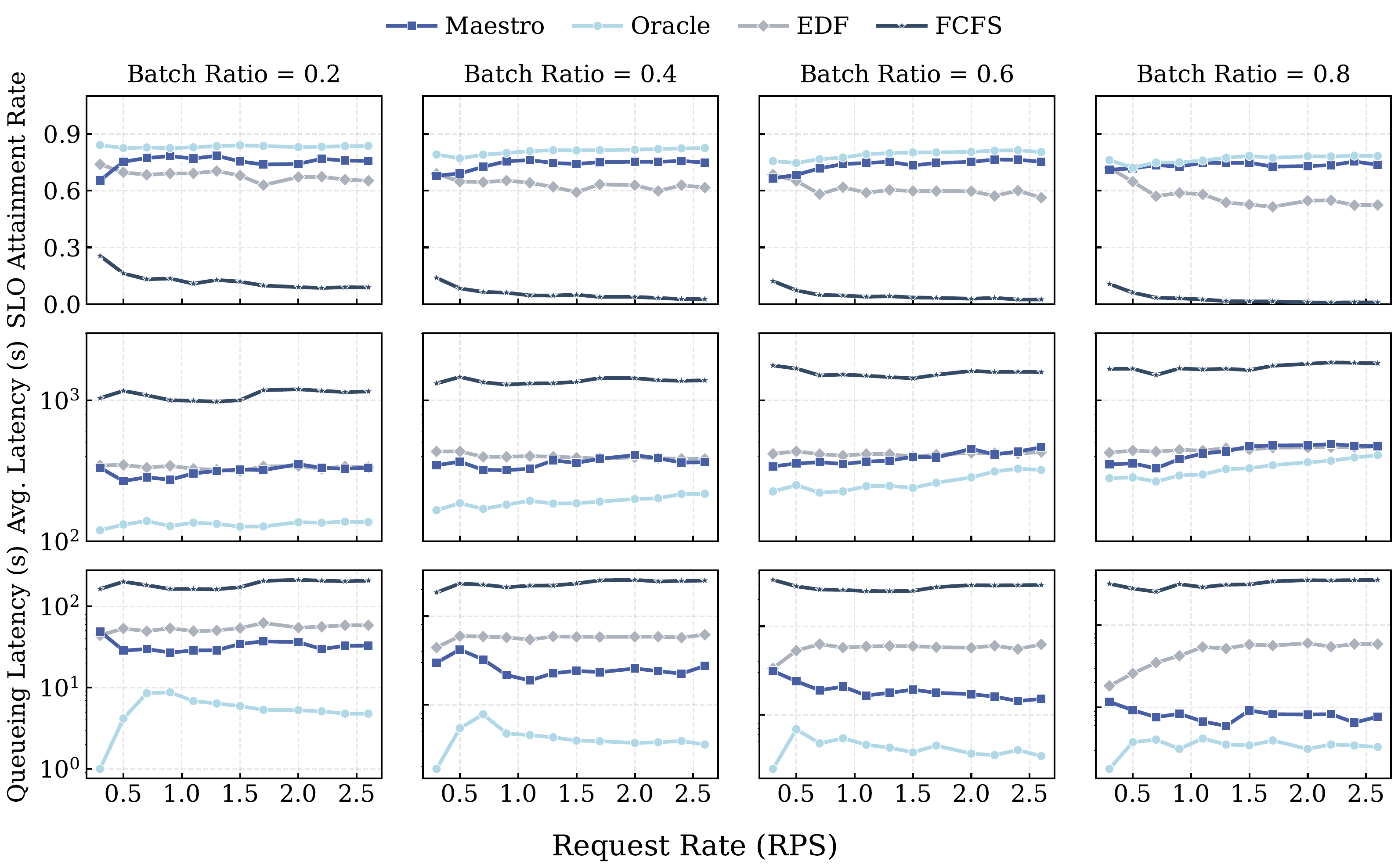}
  \caption{Overall scheduling results across arrival rates and batch ratios: SLO attainment, mean latency, and interactive queueing delay.}
  \label{fig:simulation_results_final_layout}
\end{figure}

\begin{table}[t!]
\centering
\caption{Workload trace summary (I indicates Interactive, B indicates Batch).}
\label{tab:workload_trace}
\small
\resizebox{\linewidth}{!}{
\begin{tabular}{lccc}
\toprule
App. & Type & \#Jobs & Input Source \\
\midrule
Meeting Booking & I & 8626 & Synthetic \\
Document Writing & B & 8319 & Synthetic \\
News Collection & B & 6616 & Synthetic \\
Performance & B & 6548 & IBM HR Analytics~\cite{ibm-hr-analytics-employee-attrition-performance} \\
QA Assistant & I & 5849 & MATH~\cite{hendrycksMeasuringMathematicalProblem2021}/PM~\cite{baiTrainingHelpfulHarmless2022}/GPQA~\cite{reinGPQAGraduateLevelGoogleProof2023}/SQuAD~\cite{rajpurkarSQuAD100000Questions2016} \\
Text Translation & B & 5124 & OpenCSG Chinese Corpus~\cite{yu2025opencsgchinesecorpusseries} \\
Food Assistant & I & 3334 & Synthetic \\
Travel Assistant & I & 1543 & Synthetic \\
Code Refactoring & B & 810 & LongBench v2~\cite{bai2025longbenchv2deeperunderstanding} \\
\midrule
\textbf{Total} & -- & \textbf{46,769} & \textbf{144,524 stages} \\
\bottomrule
\end{tabular}}
\end{table}

\cpara{Baselines.}
We compare \projtitle against the following baselines.

i) \textit{Prediction}.
We compare our two-stage \textit{\projtitle-Pred} (tool-intent + length regression) with:
a) \textit{Linear} (prompt-length-only regression);
b) \textit{BERT-MLP}~\cite{qiuEfficientInteractiveLLM2024} (semantic embedding + MLP, single-stage);
and c) \textit{Magnus}~\cite{chengEnablingEfficientBatch2024} (semantic embedding + regression). We train the predictor on recorded stage-level traces and evaluate it using a stratified temporal holdout: within each agent, tool-use, and thinking-mode group, the latest 20\% of records are used as the test set and earlier records are used for training. For LightGBM regressors, the training records are further split by the same strata, using the latest 15\% as a validation set for early stopping; all reported prediction numbers are measured on the held-out test records.

ii) \textit{Scheduling}.
We implement:
a) \textit{FCFS} (global FIFO);
b) \textit{EDF} (deadline-first for batch jobs, class-priority for interactive stages);
and c) \textit{Oracle-SRTF} (shortest true remaining time with perfect knowledge), as an upper bound. All scheduling baselines share the same vLLM backend, measured model profiles, dynamic batching support, and workload arrivals; they differ only in admission, routing, and queue ordering.

iii) \textit{Multi-model serving}.
We compare against:
(a) \textit{No-Colocation} (one model per GPU);
and (b) \textit{QLM-style Switching}~\cite{patke2024queue} (multi-model with process-level restart/reload). We use QLM-style switching because it is the closest public baseline for multi-model SLO-oriented serving; No-Colocation represents the fully warm upper-cost configuration, while \projtitle evaluates the benefit of retaining warm contexts without dedicating one GPU per model.

\cpara{SLOs.}
We evaluate mixed SLOs consistent with agentic applications. For each workflow template, we first profile isolated executions and use the observed completion-time distribution to set workload-specific deadlines; batch-style jobs use end-to-end completion deadlines, while interactive jobs emphasize responsiveness. Since scheduling primarily affects time-to-start (queueing + activation) rather than per-token decode speed, we approximate user-perceived delay by the sum of per-stage waiting times, and report both interactive queueing delay and SLO attainment accordingly. The same SLO targets are used for all baselines.

\subsection{Overall Performance}
\label{sec:eval_overall}

We first evaluate end-to-end performance under mixed-SLO workloads using trace-driven simulation. The request arrival rate is varied from low load ({\inlinemathsize $\lambda=0.4$} req/s) to high load ({\inlinemathsize $\lambda=2.0$} req/s), and the batch ratio from 0.20 to 0.80.
Figure~\ref{fig:simulation_results_final_layout} reports SLO attainment, mean latency, and interactive queueing delay.

Under high load and large batch ratios, \textit{FCFS} collapses due to severe head-of-line blocking, reducing SLO attainment to below 10\%. \textit{EDF} improves performance by prioritizing imminent deadlines but degrades when long jobs approach deadlines, causing short interactive stages to be delayed. In a representative high-stress configuration ({\inlinemathsize $\lambda=2.0$}, batch ratio=0.8), EDF achieves 50.0\% SLO attainment, whereas \projtitle achieves 73.6\%, corresponding to a 23.6 percentage points improvement. \projtitle further reduces interactive queueing delay by 84.8\% relative to EDF, highlighting the effectiveness of remaining-time-aware prioritization.

\begin{table}[t]
  \centering
  \caption{Ablation study of the Preemption in \projtitle}
  \label{tab:topology_impact}
  \begin{tabular}{lccccc}
    \toprule
    \multirow{2}{*}{\#Nodes} & \multicolumn{2}{c}{\textbf{\projtitle{}}} & & \multicolumn{2}{c}{\projtitle{} w/o Preempt} \\
    \cmidrule{2-3} \cmidrule{5-6}
    & SLO $\uparrow$ & Delay (ms) $\downarrow$ & & SLO $\uparrow$ & Delay (ms) $\downarrow$ \\
    \midrule
    1 & 29\% & 538k & & 29\% & 1500k \\
    2 & \textbf{60\%} & \textbf{389} & & 27\% & 235k \\
    3 & \textbf{38\%} & \textbf{22} & & 26\% & 82k \\
    4 & \textbf{75\%} & \textbf{9} & & 59\% & 19k \\
    5 & \textbf{85\%} & \textbf{2} & & 84\% & 11k \\
    \bottomrule
  \end{tabular}
\end{table}

We further evaluate the impact of preemption under extreme load ({\inlinemathsize $\lambda=5.0$}, batch ratio=0.6) by comparing \projtitle with and without preemption. Table~\ref{tab:topology_impact} reports SLO attainment and interactive queueing delay as the number of nodes increases. Preemption mitigates head-of-line blocking from long-running batch stages: it raises SLO attainment on two and four nodes (60\% vs.\ 27\%; 75\% vs.\ 59\%), and still cuts interactive queueing delay by orders of magnitude even when SLOs align (2\,ms vs.\ 11\,s on five nodes). These results use boundary preemption (between LLM invocations); the common overhead is limited to priority checking and re-queueing, while resume costs from reclaimed state are reflected in the scheduler's degradation-cost term (Eq.~\ref{eq:deg_cost}).

\begin{table}[t]
  \centering
  \caption{Tool-intent classification comparison ($\uparrow$ higher is better; $\downarrow$ lower is better).}
  \label{tab:tool_call_model_comparison}
  \footnotesize
  \setlength{\tabcolsep}{4pt}
  \renewcommand{\arraystretch}{1.15}

  \begin{tabular}{lccc}
    \toprule
    \multirow{2}{*}{Model} & AUC $\uparrow$ & F1 Macro $\uparrow$ & Acc. $\uparrow$ \\
                           & MSE $\downarrow$ & LogLoss $\downarrow$ & False Recall $\uparrow$ \\
    \midrule

    \multirow{2}{*}{\textbf{\projtitle{}-Pred}}
      & \textbf{0.9625} & 0.8999 & 0.8999 \\
      & \textbf{0.0728} & \textbf{0.2437} & 0.8966 \\
    \midrule

    \multirow{2}{*}{MLP\_64\_32}
      & 0.9535 & 0.8838 & 0.8840 \\
      & 0.0827 & 0.2837 & \textbf{0.9115} \\
    \midrule

    \multirow{2}{*}{MLP\_128\_64}
      & 0.9520 & 0.8802 & 0.8802 \\
      & 0.0834 & 0.2948 & 0.8906 \\
    \midrule

    \multirow{2}{*}{MLP\_BayesOpt\_3}
      & 0.9609 & \textbf{0.9022} & \textbf{0.9022} \\
      & 0.0736 & 0.3750 & 0.8981 \\
    \midrule

    \multirow{2}{*}{ResNet}
      & 0.9484 & 0.8838 & 0.8840 \\
      & 0.0909 & 0.3327 & \textbf{0.9115} \\
    \midrule

    \multirow{2}{*}{TabTransformer}
      & 0.9502 & 0.8848 & 0.8848 \\
      & 0.0868 & 0.2882 & 0.8801 \\
    \bottomrule
  \end{tabular}
  \vspace{-1.5em}
\end{table}

\subsection{Resource Efficiency}
\label{sec:eval_efficiency}

We evaluate the impact of multi-model colocation on GPU efficiency and responsiveness using (i) replay of real multi-model workflows and (ii) node-level utilization and memory-accounting analysis.

\subsubsection{Cost-latency trade-off under tight GPU budgets}
\label{sec:eval_cost_latency}

We evaluate application-level performance using a real multi-model \textit{Travel Assistant} workflow comprising six LLM invocations across three models. Table~\ref{tab:latency_comparison} reports end-to-end completion time under varying GPU budgets. With three GPUs, both approaches match the performance of exclusive deployment. As the GPU budget decreases, QLM incurs substantial cold-start overheads, whereas \projtitle maintains responsiveness, reducing completion time by 38.9\% with two GPUs and 70.0\% with one GPU. Relative to exclusive deployment, \projtitle incurs modest completion-time increases (3.0\% with two GPUs and 12.1\% with one GPU) while achieving significant GPU cost reductions (33.3\% and 66.7\%, respectively), demonstrating a favorable cost-latency trade-off.

\begin{table}[t]
  \centering
  \caption{Travel Assistant completion time vs.\ GPU budget.}
  \label{tab:latency_comparison}
  \begin{tabular}{lccc}
    \toprule
    Method & 1 GPU (s) $\downarrow$ & 2 GPUs (s) $\downarrow$ & 3 GPUs (s) $\downarrow$ \\
    \midrule
    \textbf{\projtitle{}} & \textbf{92.4} & \textbf{84.9} & 82.4 \\
    QLM & 309.1 & 139.1 & 82.4 \\
    \bottomrule
  \end{tabular}
\end{table}

\subsubsection{Utilization and memory overcommitment under colocation}
\label{sec:eval_overcommit}

\begin{figure}[t]
  \centering
  \includegraphics[width=\linewidth]{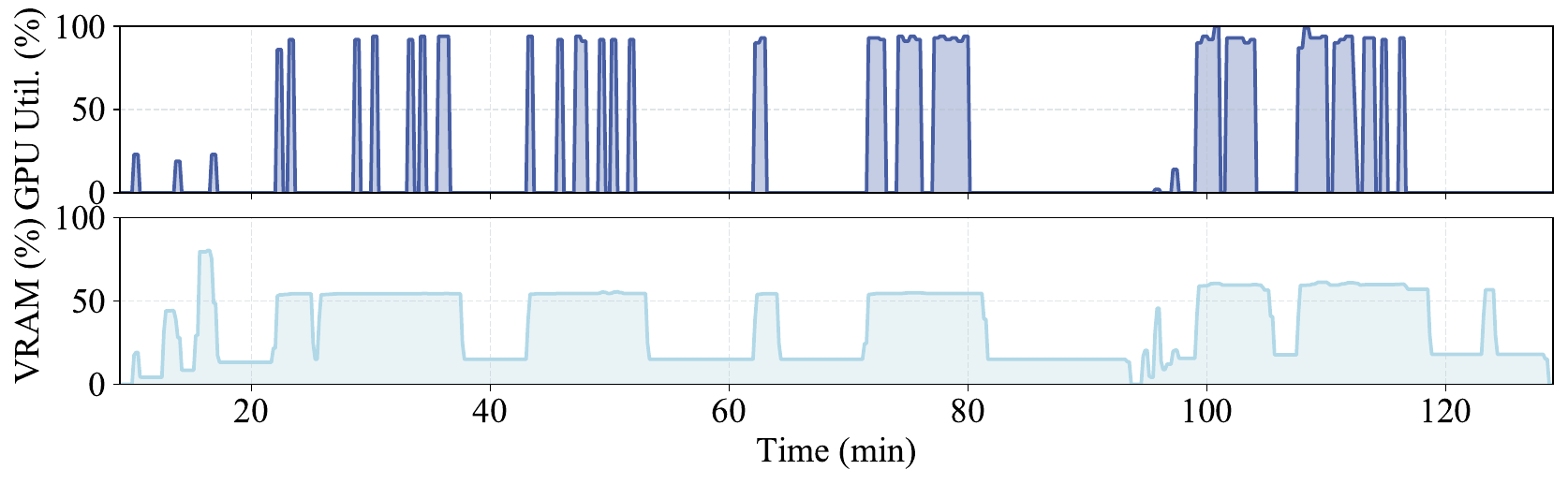}
  \caption{GPU utilization and memory usage under multi-model colocation.}
  \label{fig:GPU_Stacked_Plot}
\end{figure}

Figure~\ref{fig:GPU_Stacked_Plot} illustrates GPU utilization and memory usage when multiple Qwen3 models are colocated on a single A100 GPU under low load. GPU utilization remains highly intermittent, indicating that exclusive deployment would waste substantial capacity. In contrast, \projtitle maintains models in warm states and safely overcommits memory, enabling rapid response without dedicating separate GPUs to each model.

Table~\ref{tab:model_memory_details} reports runtime memory accounting for colocating five models. The total virtual KV-cache budget reaches approximately 122GB on a 40GB GPU, corresponding to a 
3.05× effective memory footprint, i.e., 205\% overcommitment or 67.2\% less HBM than
non-overcommitted reservation. The preserved CUDA-graph/runtime contexts consume 194--286MB per model (about 1.15GB in total for this five-model configuration), so the sleeping mechanism is not free; this footprint is explicitly counted in $M_{\mathrm{res}}$ and traded against activation latency. This level of overcommitment is feasible because actual KV usage is elastic and admission is guided by predicted demand, enabling substantial cost savings in low-load, long-tail multi-model scenarios.

\begin{table}[t]
  \centering
  \caption{Memory accounting under five-model colocation on a single A100 40GB GPU.}
  \label{tab:model_memory_details}
  \begin{tabular}{lccc}
    \toprule
    Model & CUDA Graph & Weight Size & Virtual KV Cache \\
    \midrule
    Qwen3-0.6B & 194 MB & 1.12 GB & 34.24 GB \\
    Qwen3-1.7B & 194 MB & 3.21 GB & 32.15 GB \\
    Qwen3-4B & 256 MB & 7.55 GB & 27.81 GB \\
    Qwen3-8B & 245 MB & 15.27 GB & 20.07 GB \\
    Qwen3-14B & 286 MB & 27.52 GB & 7.74 GB \\
    \bottomrule
  \end{tabular}
\end{table}

\subsection{Component Analysis}
\label{sec:eval_components}

We conclude by analyzing the key system components that contribute to the end-to-end performance.

\subsubsection{Tool-calling intent classification}
Tool-invoking stages form a distinct execution mode with short, structured outputs in agent workflows, motivating explicit intent prediction. We compare the first-stage classifier of \projtitle-Pred (LightGBM with combined semantic and structured features) against neural baselines, including MLP variants and dual-tower/Transformer-style fusion models. Table~\ref{tab:tool_call_model_comparison} reports evaluation metrics including AUC, macro-F1, accuracy, MSE, log loss, and negative-class recall.

\textit{\projtitle-Pred} achieves the highest AUC (0.9625) and the lowest MSE (0.0728) and log loss (0.2437); the three-layer MLP gains marginally on F1/accuracy but at much higher log loss, i.e., worse calibration---a decisive property because the classifier output feeds the second-stage regressor as a continuous feature.

\subsubsection{Output-length regression and ablation}
Table~\ref{tab:predict_accuracy} reports output-length prediction accuracy. \textit{\projtitle-Pred} achieves an MAE of 165.43 tokens with an {\inlinemathsize $R^2$} of 0.7774, reducing MAE by 19.2\% relative to Magnus and by 30.9\% relative to BERT-MLP. In contrast, the linear baseline yields a negative {\inlinemathsize $R^2$}, indicating that output length in multi-agent workflows is highly non-linear and cannot be captured by simple correlations with input length alone.

\begin{table}[t]
  \centering
  \caption{Output token length prediction accuracy.}
  \label{tab:predict_accuracy}
  \begin{tabular}{lcccc}
    \toprule
    Metric & \textbf{\projtitle{}-Pred} & Magnus & BERT-MLP & Linear \\
    \midrule
    MAE $\downarrow$ & \textbf{165.43} & 204.74 & 239.33 & 496.86 \\
    $R^2$ $\uparrow$ & \textbf{0.7774} & 0.6721 & 0.5620 & -0.3177 \\
    \bottomrule
  \end{tabular}
  \vspace{-1em}
\end{table}

We evaluate two ablations: (i) \projtitle-Pred w/o C removes the tool-calling intent classifier and directly regresses output length; (ii) \projtitle-Pred w/o BERT further removes semantic embeddings, relying solely on structured agent features. As shown in Table~\ref{tab:predict_ablation_study}, semantic embeddings contribute substantially, with overall {\inlinemathsize $R^2$}
 decreasing from 0.7774 to 0.7129 without BERT. Tool-calling intent classification is most beneficial in non-CoT settings, where tool-invoking and non-tool-invoking stages exhibit sharper length divergence; under CoT, this benefit diminishes as both modes include explicit reasoning segments.
 

\begin{table}[t]
  \centering
  \small
  \setlength{\tabcolsep}{3pt}
  \caption{Ablation study of the prediction module in \projtitle.}
  \label{tab:predict_ablation_study}
  \begin{tabular}{lcccc}
    \toprule
    Version & MAE $\downarrow$ & $R^2$ $\uparrow$ & MAE (CoT) $\downarrow$ & MAE (non-CoT) $\downarrow$ \\
    \midrule
    Full & \textbf{165.43} & \textbf{0.7774} & 265.97 & \textbf{134.17} \\
    w/o C & 170.78 & 0.7722 & \textbf{264.25} & 141.72 \\
    w/o BERT & 175.46 & 0.7129 & 288.18 & 140.42 \\
    \bottomrule
  \end{tabular}
\end{table}

\subsubsection{Online overhead}
\begin{figure}[t]
  \centering
  \includegraphics[width=\linewidth]{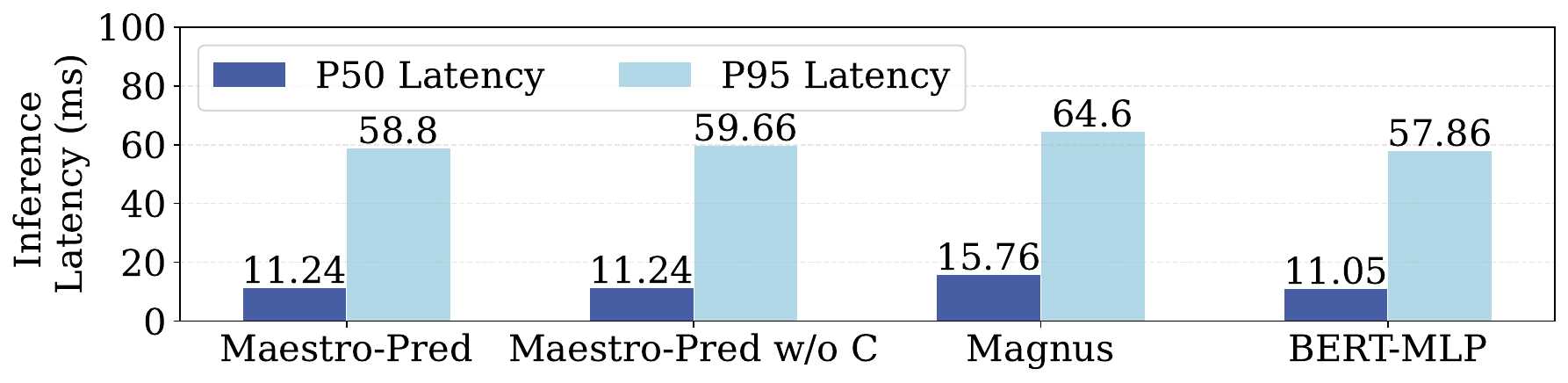}
  \caption{Prediction overhead (P50/P95) for methods with BERT encoders.}
  \label{fig:Prediction_Latency}
  \vspace{-1em}
\end{figure}

Prediction latency directly affects time-to-first-token. We measure end-to-end inference latency of the prediction module, including BERT encoding. Figure~\ref{fig:Prediction_Latency} reports P50/P95 latency for BERT-based methods. \textit{\projtitle-Pred} achieves a median latency of 11.24ms, comparable to BERT-MLP and significantly lower than Magnus (15.76ms). Adding the first-stage intent classifier incurs negligible overhead, as both stages share feature extraction and preprocessing.

\begin{table}[t]
  \centering
  \caption{Impact of node fitness scoring on interactive queueing delay.}
  \label{tab:binpack_effectiveness}
  \begin{tabular}{lccc}
    \toprule
    Arrival rate $\lambda$ & Baseline (s) & BinPack Only (s) & \projtitle{}-Aff (s) \\
    \midrule
    0.5 & 0.12 & 0.05 & \textbf{0.03} \\
    1.0 & 1.79 & 1.43 & \textbf{1.23} \\
    2.0 & 36.99 & 33.67 & \textbf{30.77} \\
    \bottomrule
  \end{tabular}
\end{table}

\subsubsection{Runtime Support for Colocation: Model Activation Latency}
\label{sec:eval_model_activation}
\begin{figure}[t]
  \centering
  \includegraphics[width=\linewidth]{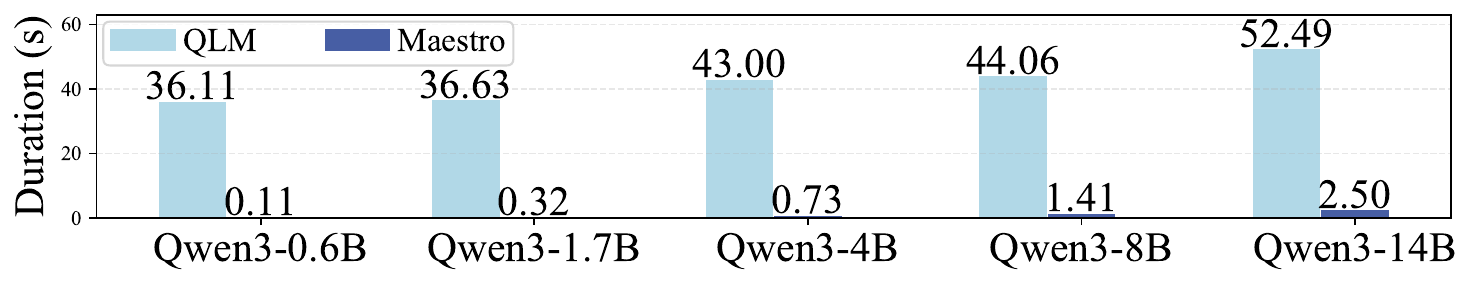}
  \caption{Model activation latency (0.6B--14B) on a single A100 40GB GPU.}
  \label{fig:Latency}
\end{figure}

Figure~\ref{fig:Latency} compares model startup and activation latency for models ranging from 0.6B to 14B parameters on a single A100 40GB GPU. Unlike QLM-style approaches that require process-level restarts for model switching~\cite{patke2024queue}, \projtitle reduces activation latency by reusing runtime contexts and avoiding process restarts, enabling practical multi-model serving under long-tail demand.

\subsubsection{Cross-Cluster Dispatch and Node Fitness Scoring}
\label{sec:eval_fitness}

To evaluate cross-cluster scheduling, we configure the simulator with a hybrid topology comprising three local and two remote nodes. The physical cluster validates node-level execution and colocation, while this simulator isolates cross-cluster routing effects using the measured model profiles and RTT matrix from Figure~\ref{fig:aliyun_latency}. We compare three policies: (i) Baseline, which performs simple load balancing without prediction-guided bin packing; (ii) BinPack Only, which enables KV-demand-aware bin packing but ignores network latency ({\inlinemathsize $\gamma=0$}); and (iii) \projtitle-Aff, which applies the full fitness score with a latency weight ({\inlinemathsize $\gamma=0.25$}) to penalize remote placements. Table~\ref{tab:binpack_effectiveness} reports the average interactive queueing delay under varying arrival rates {\inlinemathsize $\lambda$}. Fitness weights are selected on the validation split and kept fixed across test workloads; robust normalization in Eq.~\ref{eq:fitness} prevents any single metric from dominating under outlier RTT or activation estimates.

KV-demand-aware bin packing substantially reduces queueing delay by mitigating fragmentation and consolidating compatible requests, achieving a 58.3\% reduction relative to Baseline under low load ({\inlinemathsize $\lambda=0.5$}) and remaining effective under high load ({\inlinemathsize $\lambda=2.0$}).
Incorporating network latency in \projtitle-Aff further improves performance by favoring local execution for latency-sensitive stages, demonstrating the effectiveness of prediction-guided cross-cluster scheduling.

\section{Related Work}
\label{sec:related_work}

\cpara{Agentic workflows and system implications.}
LLM-MAS systems have evolved from ReAct-style loops~\cite{yao2022react} to plan-based decompositions~\cite{wang2023plan} and mixtures of specialized experts~\cite{wang2024mixture,guoLargeLanguageModel2024}, introducing structured heterogeneity and dependency amplification that place unique demands on serving infrastructure.
Parrot~\cite{linParrotEfficientServing2024} exposes semantic variables to optimize dataflow and Hermes~\cite{liu2025efficientservingllmapplications} models demand uncertainty via probabilistic graphs and Gittins-index policies; \projtitle further introduces a tool-intent-aware cost estimator driving SRTF to attack dependency-induced head-of-line blocking.

\cpara{Efficient serving and resource management.}
Modern LLM serving engines, such as vLLM~\cite{kwonEfficientMemoryManagement2023} and SGLang~\cite{zhengSGLangEfficientExecution2024}, improve throughput through paged attention and structured execution. MuxServe~\cite{duanMuxServeFlexibleSpatialtemporal2024} and Castor~\cite{luo2025castor} further increase GPU utilization through spatial-temporal multiplexing. Others target architecture-specific optimizations: UnifiedServe~\cite{zhao2025enablingdisaggregatedmultistagemllm} optimizes multi-stage MLLM inference via GPU resource sharing, FinDEP~\cite{pan2025efficientmoeinferencefinegrained} improves MoE inference via fine-grained expert scheduling, and Cauchy~\cite{zhang2025cauchy} adaptively places prefill and decode on heterogeneous GPUs. Similar orchestration principles appear in adjacent domains, including OREO~\cite{MungariO-RAN2025} for O-RAN and Mina~\cite{Mina2025} for in-network aggregation.
Serverless and elastic frameworks such as Torpor~\cite{YuTorpor2025}, SMLT~\cite{ali2025enabling}, Espresso~\cite{Espresso2025}, and KAIOps~\cite{kaiopas2025ase} address high concurrency and heterogeneous GPU resources, while QLM~\cite{patke2024queue} and SELA~\cite{tuli2025sela} manage model switching to meet SLOs. However, these systems largely treat requests as independent. In contrast, \projtitle incorporates agent-level context and workflow dependencies into memory coordination and scheduling, enabling efficient multi-agent orchestration.

\cpara{Scheduling and performance prediction.}
Previous work predicts LLM output length using bucketed models (S$^3$~\cite{jinS3IncreasingGPU2023}), semantic embeddings (Magnus~\cite{chengEnablingEfficientBatch2024}), or neural regressors~\cite{qiuEfficientInteractiveLLM2024}, but typically ignores the structured context of multi-turn, role-driven agent workflows. In GPU cluster scheduling, Kale~\cite{kale2024socc} leverages traffic forecasting for elastic autoscaling, Cuckoo~\cite{cuckoo2025socc} jointly optimizes deadline satisfaction and utilization via spatio-temporal packing, A-SRPT~\cite{luo2025predictionassistedonlinedistributeddeep} couples prediction with SRPT scheduling, and KAIR~\cite{kair2025ase} employs causal inference to diagnose training stragglers. \projtitle extends prediction-driven scheduling to the multi-agent setting by combining a two-phase tool-intent predictor with fitness-based cross-cluster routing and workflow-aware remaining-time queueing.
\section{Conclusion}
\label{sec:conclusion}

We presented \projtitle, a workload-aware scheduling system for LLM-MAS. Rather than treating LLM requests as opaque, \projtitle characterizes the workload through agent-level dimensions---role and workflow position, tool-invocation intent, predicted per-stage output length and KV footprint, and per-cluster model readiness---and drives hierarchical weight residency, elastic KV provisioning, fitness-based dispatch, and boundary-preemptive SRTF in a single control loop. 
Across prototype experiments and trace-driven simulations, \projtitle
reduces KV-reservation HBM by 67.2\% and improves high-contention SLO
attainment over EDF by 23.6 percentage points.

\noindent\textbf{Limitations and future work.}
\projtitle complements---rather than replaces---kernel-/model-level optimizations (speculative decoding, KV compression, adapter-based serving, disaggregated prefill/decode, iteration-level schedulers); these can be folded in by updating per-model microbenchmarks and readiness signals. Our current prototype focuses on homogeneous A100 nodes and two service classes; production deployments with heterogeneous accelerators and multiple tenants require hardware-aware placement constraints, quota enforcement, and stronger fairness guarantees. Finally, although tool latency is recorded in workflow profiles when available, jointly scheduling external tool execution with LLM inference remains an important direction for tool-heavy agent systems.


\section*{Acknowledgment}
This work was supported in part by National Key R\&D Program of China (Grant No. 2024YFB4505604), in part by the National Natural Science Foundation of China (Grant No. 62402024), in part by the Fundamental Research Funds for the Central Universities. 


\printbibliography

\end{document}